\documentclass[a4paper, 12pt]{article}
\usepackage{cite}
\usepackage{rotating}
\usepackage[english]{babel}
\usepackage[a4paper, inner=2cm, outer=2cm, top=3cm, bottom=3cm]{geometry}
\usepackage[dvipsnames]{xcolor}
\usepackage{mathtools}
\usepackage{pstricks}
\usepackage{caption}
\usepackage{graphicx}
\usepackage{amsmath}
\usepackage{amssymb}
\usepackage{mathcomp}
\usepackage{textcomp}
\usepackage{enumitem}
\usepackage{physics}
\usepackage{romannum}
\usepackage[doublespacing]{setspace}
\usepackage{subcaption}
\usepackage{hyperref}

\hypersetup{
  colorlinks   = true,
  urlcolor     = Blue, 
  linkcolor    = Black,
  citecolor    = Black
}

\renewcommand{\thesection}{\Roman{section}} 

\title{On the quantum gravastar}
\author{Raihaneh Moti$^{1}$ and Ali Shojai$^{2}$\\
$^{1}$\textit{\small Department of Physics, Ferdowsi University of Mashhad, Mashhad, Iran}\\
$^{2}$\textit{\small Department of Physics, University of Tehran, Tehran, Iran}}
\date{}

\begin{document}
\maketitle
\pagenumbering{arabic}
\begin{abstract}
In this paper, we show that it is possible to have pure quantum gravastar solution to the quantum improved Einstein equations. Such an object is essentially a gravastar in which the thin layer of matter is replaced by quantum effects. The metric of a pure quantum gravastar is obtained and its stability is discussed.
\end{abstract}

\section{Introduction}
Simply stated, a black hole is a gravitational collapsing core, surrounded by an event horizon. The causal structure may undergo notable changes by passing the null boundary surface (event horizon) defined as where the light cone closes\cite{Horizon}.
Although the main features of a black hole are attributed to the event horizon, but some non--analytical aspects of this surface besides the core singularity, diverted some attentions to alternative models. The \textit{Mazur--Mottolla gravastar} mainly stands for this idea\cite{Mazur}.

On their suggestion, replacement of the black hole collapsing matter with a deSitter--like core which is surrounded by a layer of matter of finite thickness, prevents gravitational collapse and infinite density singularity formation. The thin layer is composed of a positive pressure fluid with typical ultra relativistic equation of state $P = + \rho$. Thus, the gravastar's three main regions are internal deSitter core, matter layer with a positive pressure and the external vacuum region, which for spherically symmetric static spacetime is necessarily in Schwarszchild form according to the Birkhoff's theorem.

To have a well--defined smooth manifold, the fundamental forms should not experience any discontinuity at the boundaries of these regions, unless some Darmois--Israel matter compensates them\cite{Israel, Poisson}. As a result, the intermediate thin layer practically consists of three sub--layers: An external thin shell at radius $r_{e} \gtrsim r_{H_e}$ (where $r_{H_e}$ is the horizon of the external region) with surface density $\sigma_{e}$ and surface tension $\vartheta_{e}$ , a thin layer which is filled with a positive pressure fluid, and an internal thin shell at  $r_{i} \lesssim r_{H_i}$  (where $r_{H_i}$ is the horizon of the internal region) with surface density $\sigma_{i}$ and surface tension $\vartheta_{i}$.

Clearly, no singularity is present in the deSitter core. Further, the presence of junction layers, solves the problem of non--global time coordinate. Moreover, with some exclusive considerations, this object do not experience any thermodynamical unstability\cite{Mazur}.

The gravastar idea is based on the Chapline--Hohlfled--Laughlin--Santiago model\cite{Chapline}. They proposed to replace the event horizon with a critical phase transition layer where the Bose--Einstein condensation occurs in a gravitational system, so that the classical general relativity remains valid arbitrary close to the horizon. It is quite possible that the quantum effects are  dominant  at the horizon and  the general theory of relativity  breaks there. 

As a result, it would be notable if we look for the replacement of gravastar thin layer with one of quantum nature, which holds the quantum effects of geometry, \textit{a quantum improved layer}. 

An improved gravitational system contains corrections which raised from considering the effects of the renormalization group corrections. This is based on using the exact renormalization group equation to study the Weinberg asymptotic safety conjecture. According to this renormalization method, any quantum field theory is UV complete if its essential couplings run to a non--Gaussian fixed point at UV\cite{AS1,AS2}. The existence of such a fixed point for the gravity renormalization group flow, is searched by various methods (see references in \cite{Weinberg-inf}).

The effects of the quantization are usually studied using the effective average action. Unfortunately, solving the exact renormalization group equation to derive the effective average action is complex, if not impossible. Therefore, by the decoupling idea \cite{Reuter & Saueressig, Reuter & Weyer} these effects are considered as some sort of correction to the classical theory and studied through an effective theory obtained by improving the classical coupling constant to the running one which is derived from the solution for the $\beta$-function\cite{Reuter-1st,Reuter & Weyer,Rodrigues,Reuter-3rd}. This can be done with different strategies as it is discussed in\cite{our3}. The most physical way of quantum improvement of the field equations, is the action improvement presented in\cite{our3}. 

To simplify the model, usually the theory space is truncated to Einstein--Hilbert subspace and the whole effective average action is expanded using two basis: $\sqrt{g}$ and $\sqrt{g} R$ \cite{Codello}. This truncation has satisfactory results, especially when the desirable spacetime is the vacuum one. As a result of the Einstein--Hilbert truncation, the renormalization group equation leads to
\begin{equation} \label{RC}
G(\chi) = \frac{G_0}{1+f(\chi)}
\end{equation}
where $f(\chi) \equiv \xi /\chi^2$ is a dimensionless function of the renormalization group parameter $\chi$. This parameter has a dimension of length as one expects. The small scaling constant $\xi$ (length square dimensional) equals to $ \omega_q \xi_0 G_0$  provided that the reference constant $ G_0 $ be the experimentally observed value of Newton's constant $G_N$ and $\omega_q = \frac{4}{\pi}(1-\frac{\pi^{2}}{144})$ is the quantization parameter obtained from the renormalization group methods\cite{Bonanno & Reuter}. The parameter $\xi_0 $ (of order one) is used to relate the length and mass dimensions, $[M] = \xi_0 [L]^{-1}$ .

The renormalization group parameter $\chi$ is the scaling parameter of the theory. It is discussed in detail in \cite{our3} that the best scaling parameters of the spacetime are the components of the Riemann tensor, which physically describe the tidal forces. To save the general covariance of the action after improvement, they should contribute in $\chi$ in terms of curvature invariants. Thus, in general, $\chi$ could be a well--defined function of all independent curvature invariants such as $R, R_{\alpha\beta}R^{\alpha\beta}, R_{\alpha\beta\gamma\delta}R^{\alpha\beta\gamma\delta}, \ldots $.

Unfortunately, one of the unwanted features of this method is that there is no unique way to fix the form of $\chi$, in general yet\cite{Pawlowski,Reuter-2021}. Considering various conditions such as the behavior of singularities and energy conditions\cite{our4, our5}, one can restrict the possible choices, but not fix them universally.

Accordingly, it seems a plausible idea to have a region of deSitter spacetime confined by a thin quantum (gravitational) layer, instead of a thin layer of matter. This is what we are looking for in this paper and call such an object a \textit{pure quantum gravastar}. To have such a solution, the corrections from the improved Einstein--Hilbert action, with a proper cutoff function $f(\chi)$ should dominate about the classical horizons.
A pure quantum gravastar is consisted of a deSitter core and a Schwartzschild outer region for which quantum gravitational effects are negligible for both of them. These two regions are connected smoothly to each other by a thin layer of quantum gravitational effects. In this way one has spacetime effects similar to black holes, without any matter.

\section{Quantum gravastar}
Here we are looking for the pure quantum gravastar solution, a vacuum (no matter, but quantum gravitational effects) solution of quantum improved Einstein equations.

\subsection{Kinematic description}
By a pure quantum vacuum solution of quantum improved Einstein's equations, we mean a solution with a deSitter core and an external Schwartzchild region. These two regions should be matched to each other by a thin layer of vacuum full of quantum effects of gravity.
  
Such a substitution of the thin fluid layer of gravastar with a quantum vacuum layer is possible if it matches smoothly with both internal and external sides, and all the internal, external and the layer metrics be solutions of the quantum improved equations. 

To compare the results with Mazur--Mottola gravastar, we consider the general spherical symmetric metric
\begin{equation} \label{SSM}
ds^2 = f(r)\dd{t}^2 -h(r)^{-1}\dd{r}^2-r^2 \dd{\Omega}_2
\end{equation}
where $\dd{\Omega}_2 \equiv \dd{\theta}^2 +  \sin\theta^2 \dd{\phi}^2 $. The deSitter core is described by
\begin{equation} \label{Internal-Solution}
f(u<u_d) \equiv f_i(u) = 1 - \tilde{H}_0^2 u^2 \quad \quad , \quad \quad  h(u<u_d) \equiv h_i(u) = 1 - \tilde{H}_0^2 u^2
\end{equation}
and the metric for the external Schwarzschild region is
\begin{equation} \label{External-Solution}
f(u>u_s) \equiv f_e(u) = K( 1 - \sigma/u) \quad \quad , \quad \quad  h(u>u_s) \equiv h_e(u) = K( 1 - \sigma/u)
\end{equation}
where $u \equiv r/l_p$, $ \sigma \equiv r_s/l_p$ and $\tilde{H}_0 \equiv H_0 l_p^2$ are the normalized counterpartes of the radial coordinate, Schwarzschild radius and deSitter constant, respectively. 
Note that as we are looking for a pure quantum solution, it is natural to scale everything with the planck length, $l_p$. 
The  $u_d$ and $u_s$ are the boundaries of the deSitter and Schwarzscild regions. Also, $K$ is a constant needed for matching different parts of the solution.

The intermediate region $u_d<u<u_s$ can be restricted to have spherical symmetry, for comparison of the results with Mazur--Mottola gravastar. Therefore we employ the spherical symmetric solution  
\begin{equation}
ds^2 = f_q(r)\dd{t}^2 -h_q(r)^{-1}\dd{r}^2-r^2 \dd{\Omega}_2
\end{equation}
for this region.
In order to have a solution without any horizon, deSitter and Schwartzschild horizons should lie within this region.

All these metric components ($f_i, h_i, f_e, h_e, f_q, h_q$) should satisfy the quantum improved Einstein's equations and match smoothly to each other.

\subsection{Dynamic confirmation}
The quantum improved Einstein's equations are derived from the improved Einstein--Hilbert action \cite{our3,our2} 
\begin{equation} \label{IA}
\mathcal{A}^{\text{I}} = \int\dd[4]{x}  \frac{\sqrt{-g}}{8\pi G(\chi)} R  + \mathcal{A}_{\text{matter}} \ .
\end{equation}
As it is stated in the Introduction section, $\chi$ is the scaling parameter which should be fixed by various considerations.  Although all the independent curvature invariants may be necessary for scaling the spacetime, but they will involve the problem without any clear wisdom about the geometric description of each term. Using a length dimensional function of three fundamental curvature invariants $R, R_{\alpha\beta}R^{\alpha\beta}$ and $ R_{\alpha\beta\gamma\delta}R^{\alpha\beta\gamma\delta} $  makes the  comparison of the results with $f(R)$ models possible besides giving sufficient insight.

There are two notable conditions that would be useful to fix $\chi$:
\begin{itemize}
\item For an action which is a collection of Euler--densities, Darmois--Israel junction conditions are known to some extent\cite{Euler-Junction}. If we restrict our choices to the fundamental invariants $(R,  R_{\alpha\beta}, R_{\alpha\beta\gamma\delta}R^{\alpha\beta\gamma\delta})$, the running coupling \eqref{RC} suggests a proper combination of $1/R$, $ R/  R_{\alpha\beta}R^{\alpha\beta}$ and $ R/ R_{\alpha\beta\gamma\delta}R^{\alpha\beta\gamma\delta}$ for $\chi^2$.
\item On the other hand, we are interested in a solution for which the quantum effects are dominant at the intermediate layer and fall off out of it. This would help us to find the most impressive term of the linear composition $\chi^2 =\varphi_1 / R + \varphi_2 \times R/R_{\alpha\beta}R^{\alpha\beta} + \varphi_3 \times R/ R_{\alpha\beta\gamma\delta}R^{\alpha\beta\gamma\delta}$ with constant coefficients $\varphi_{i=1,2,3}$. Here we choose the second term, as it is discussed in the Appendix.
\end{itemize}
Therefore, we set $\chi^2 =R/R_{\alpha\beta}R^{\alpha\beta}$ and the field equations are now \cite{Higher Derivative}
\begin{equation} \label{IEoM}
G_{\alpha\beta} + \xi_p \mathcal{G}^{\text{(q)}}_{\alpha\beta} = \kappa_{0} T_{\alpha\beta}
\end{equation}
where $\xi_p \equiv \eval{\xi}_{G_0 \sim l_p^2} = \omega_q \xi_0 l_p^2$ and
\begin{equation} \label{Iterm}
 \mathcal{G}^{\text{(q)}}_{\alpha\beta} \equiv 2 R_{\alpha\mu\beta\nu}R^{\mu\nu} - \nabla_{\alpha}\nabla_{\beta} R + \square R_{\alpha\beta} -\frac{1}{2} g_{\alpha\beta} \Bigl( R_{\mu\nu}R^{\mu\nu} - \square R \Bigr) \ . 
\end{equation}

For a spherical symmetric solution \eqref{SSM}, the $f(u)$ and $h(u)$ are the solutions of the temporal, radial and polar coupled differential equations
\begin{align}
& G_{tt}+\alpha_q \mathcal{Q}_t=\frac{1}{u^2}\dv{}{u}[u(1-h)]  + \alpha_q \mathcal{Q}_t = 0 \label{I00-fh}\\
& G_{rr}+\alpha_q \mathcal{Q}_r= \frac{h}{uf}\dv{f}{u}+\frac{1}{u^2} (h-1)  + \alpha_q \mathcal{Q}_r = 0 \label{I11-fh} \\
& G_{\theta\theta}+\alpha_q \mathcal{Q}_\theta=\frac{1}{4f^2 u} \left( -hu(\dv{f}{u})^2 +2f^2 \dv{h}{u} +f \left( u \dv{f}{u} \dv{h}{u} +2h(\dv{h}{u} + u\dv[2]{f}{u}) \right) \right) + \alpha_q \mathcal{Q}_{\theta} = 0 \label{I22-fh}
\end{align}
where $\alpha_q \equiv \xi_p/16$. The $\mathcal{Q}_r(u), \mathcal{Q}_t(u)$ and $\mathcal{Q}_{\theta}(u)$ are nonlinear functions of $f(u)$ and $h(u)$ and their derivatives up to order 4, given by the relations:
\begin{equation}
\begin{split}
 \mathcal{Q}_t =  \frac{-1}{f^5 h^3 u^4}
{}& \Bigl( 48 h^5 f^5 - 64 h^4 f^5 + 16 h^3 f^5 + 208 h^2 u^3 h'^3 f^5 + 12 h^3 u^2 h'^2 f^5 - 32 h^4 u h' f^5  \\
{}& + 16 h^4 u^2 h'' f^5 -96 h^3 u^3 h' h'' f^5 + 16 h^4 u^3 h^{(3)} f^5 + 192 h^5 f^4 - 192 h^4 f^4 - 168 u^4 h'^4 f^4    \\
{}& + 384 h^2 u^3 h'^3 f^4 + 20 h^3 u^3 f'h'^2 f^4 -64 h^5 u f' f^4 + 32 h^4 u f' f^4-64 h^4 u h' f^4 + 24 h^4 u^2 f' h' f^4   \\
{}& + 32 h^5 u^2 f'' f^4 + 12 h^3 u^4 h'^2 f'' f^4 +104 h^4 u^3 h' f'' f^4 - 32 h^4 u^2 h'' f^4+168 h u^4 h'^2 h'' f^4 \\
{}& + 56 h^4 u^3 f' h'' f^4-192 h^3 u^3 h' h'' f^4+4 h^3 u^4 f' h' h'' f^4 +16 h^4 u^4 f'' h'' f^4 + 48 h^5 u^3 f^{(3)} f^4   \\
{}& +32 h^4 u^3 h^{(3)} f^4 + 4 h^4 u^4 f' h^{(3)} f^4 - 28 h^2 u^4 h' h^{(3)} f^4 + 8 h^5 u^4 f^{(4)} f^4 -240 u^4 h'^4 f^3  \\
{}& - 20 h^5 u^2f'^2 f^3  - 7 h^3 u^4 f'^2 h'^2 f^3 - 20 h^5 u^4 f''^2 f^3 + 64 h^5 u f' f^3-72 h^4 u^3 f'^2 h' f^3  \\
{}&-64 h^4 u^2 f' h' f^3   -  64 h^5 u^2 f'' f^3 - 120 h^5 u^3 f' f'' f^3 + 64 h^4 u^3 h' f'' f^3  - 60 h^4 u^4 f' h' f'' f^3  \\
{}&  - 12 h^4 u^4 f'^2 h'' f^3  +240 h u^4 h'^2 h'' f^3  + 32 h^4 u^3 f' h'' f^3+32 h^4 u^4 f'' h'' f^3 +32 h^5 u^3 f^{(3)} f^3 \\
{}&  -24 h^5 u^4 f' f^{(3)} f^3  + 8 h^4 u^4 f' h^{(3)} f^3 -40 h^2 u^4 h' h^{(3)} f^3 + 16 h^5 u^4 f^{(4)} f^3 + 52 h^5 u^3 f'^3 f^2  \\
{}&- 208 h^2 u^3 h'^3 f^2 + 64 h^5 u^2 f'^2 f^2   - 32 h^5 u^4 f''^2 f^2 + 30 h^4 u^4 f'^3 h' f^2 - 64 h^4 u^3 f'^2 h' f^2 \\
{}&  + 56 h^5 u^4 f'^2 f'' f^2 - 96 h^5 u^3 f' f'' f^2   - 88 h^4 u^4 f' h' f'' f^2 -24 h^4 u^4 f'^2 h'' f^2  + 104 h^3 u^3 h' h'' f^2  \\
{}&  - 48 h^5 u^4 f' f^{(3)} f^2 - 23 h^5 u^4 f'^4 f + 64 h^5 u^3 f'^3 f - 384 h^2 u^3 h'^3 f +48 h^4 u^4 f'^3 h'f  \\
{}& + 112 h^5 u^4 f'^2 f'' f + 192 h^3 u^3 h' h''f  - 48 h^5 u^4 f'^4  \Bigr)\ ,
\end{split}
\end{equation}
\begin{equation}
\begin{split}
\mathcal{Q}_r = \frac{1}{f^4 h u}
{}& \Bigl( 304 f^4 h^3 - 16 f^4 h^2 u^2 h'' - 320 f^4 h^2 + 40 f^4 u^3 h'^3 + 4 f^4 h u^2 h'^2 + 32 f^4 h u h'+64 f^4 h^2 h^{(3)} u^3  \\
{}&  +32 f^4 h u^3 h' h'' - 96 f^4 h^2 u h' + 16 f^4 h+80 f^{(3)} f^3 h^3 u^3 + 96 f^{(3)} f^3 h^2 u^4 h'  - 60 f^2 h^3 u^4 f''^2 \\
{}& - 88 h^3 u^4f'^4 +124 f h^3 u^3 f'^3 + 103 f h^2 u^4 f'^3 h' + 32 f^{(4)} f^3 h^3 u^4 + 32 f^4 h^4 u^2 f''  \\
{}&  +64 f^4 h^4 u f' + 16 f^4 h^3 u^2 f' h' -64 f^3 h^3 u^2 f'' + 64 f^3 h^2 u^4 f'' h'' + 24 f^3 h u^4 f''h'^2  \\
{}& + 176 f^3 h^2 u^3 f'' h'  - 16 f^3 h^4 u^2 f'^2 + 96 f^3 h^3 u f' +80 f^3 h^2 u^3 f' h'' + 28 f^3 h u^3 f' h'^2  \\
{}& + 16 f^3 h^2 h^{(3)} u^4 f' +  8 f^3 h u^4 f' h' h'' - 88 f^3 h^2 u^2 f' h' + 206 f h^3 u^4 f'^2 f''-192 f^2 h^3 u^3 f' f''\\
{}& +68 f^2 h^3 u^2 f'^2 -44 f^2 h^2 u^4 f'^2 h''-15 f^2 h u^4 f'^2 h'^2-192 f^2 h^2 u^4 f' f'' h' -148 f^2 h^2 u^3 f'^2 h' \\
{}& -88 f^{(3)} f^2 h^3 u^4 f' \Bigr)\ ,
\end{split}
\end{equation}
\begin{equation}
\begin{split}
\mathcal{Q}_{\theta} = \frac{-1}{f^4 h u^4} 
{}& \Bigl( 16 f^4 h^3 - 64 f^4 h^2 u^2 h'' - 24 f^4 u^4 h'^3 -24 f^4 h u^2 h'^2 - 4 f^4 h^2 h^{(3)} u^4 - 24 f^4 h^2 h^{(3)} u^3 \\
{}&  + 24 f^4 h u^4 h' h'' - 12 f^4 h u^3 h' h'' + 48 f^4 h^2 u h' - 16 f^4 h-56 f^{(3)} f^3 h^3 u^3 - 24 f^{(3)} f^3 h^2 u^4 h'  \\ 
{}&   + 12 f^2 h^3 u^4 f''^2 + 19 h^3 u^4 f'^4 - 68 f h^3 u^3 f'^3 - 22 f h^2 u^4 f'^3 h' - 8 f^{(4)} f^3 h^3 u^4-48 f^3 h^3 u^2 f''\\ 
{}& - 16 f^3 h^2 u^4 f'' h''-  6 f^3 h u^4 f'' h'^2 - 116 f^3 h^2 u^3 f'' h'+64 f^3 h^3 u f'-48 f^3 h^2 u^3 f' h'' \\ 
{}& -16 f^3 h^2 u f' +24 f^3 u^4 f'h'^3 - 16 f^3 h u^3 f' h'^2 - 26 f^3 h u^4 f' h' h''-36 f^3 h^2 u^2 f' h' \\
{}& -44 f h^3 u^4 f'^2 f''+  120 f^2 h^3 u^3 f' f''+ 44 f^2 h^3 u^2 f'^2+ 10 f^2 h^2 u^4 f'^2 h'' + 3 f^2 h u^4 f'^2 h'^2 \\
{}& + 42 f^2 h^2 u^4 f' f'' h'+  84 f^2 h^2 u^3 f'^2 h'+20 f^{(3)} f^2 h^3 u^4 f' \Bigr) \ .
\end{split}
\end{equation}

The solution of \eqref{I00-fh}-\eqref{I22-fh} should also satisfy the  effective conservation equation
\begin{equation}\label{IC}
\mathcal{Q}_c\equiv  -\frac{1}{\kappa_0} \left(\dv{\mathcal{Q}_r}{u}+(\mathcal{Q}_r+\mathcal{Q}_t)\frac{1}{2 f}\dv{f}{u} + 2 \frac{\mathcal{Q}_r-\mathcal{Q}_{\theta}}{u} \right)=0 \ 
\end{equation} 
obtained by taking the divergence of field equations \eqref{IEoM}.

It is quite easy to check that the quantum effects, $\mathcal{Q}_t$, $\mathcal{Q}_r$, $\mathcal{Q}_\theta$, and $\mathcal{Q}_c$ rapidly go to zero in the internal ($u<u_d$) and external ($u>u_s$) regions and thus the solutions \eqref{Internal-Solution} and \eqref{External-Solution} are valid. For the intermediate region we propose the Gaussian functions
\begin{equation} \label{fjhj}
f_q(u) = c_1 + \frac{a_1}{1+b_1(u/\sigma-1)^2} \quad \quad , \quad \quad  h_q(u) = c_2 + \frac{a_2}{1+b_2(u/\sigma-1)^2}
\end{equation}
with six constants $(a_1, a_2, b_1, b_2, c_1, c_2)$ to be fixed by matching conditions.

For deSitter region the estimation $\tilde{H}_0 \simeq 1/\sigma $ is  fine, and thus the matching conditions
\begin{align} \label{BC}
& f_i(u_d) = f_q(u_d)  \quad \quad , \quad \quad  h_i(u_d) = h_q(u_d)   \quad \quad , \quad \quad f_q(u_d) =h_q(u_d) \nonumber \\
& f_e(u_s) = f_q(u_s)  \quad \quad , \quad \quad  h_e(u_s) = h_q(u_s)    \quad \quad , \quad \quad f_q(u_s)=h_q(u_s)
\end{align}
result in
\begin{equation} \label{var}
 a_1 = a_2 \approx -\frac{16}{3 \sigma}  \quad \quad , \quad \quad b_1 = b_2 \approx \frac{\sigma^2}{3} \quad \quad , \quad \quad c_1 = c_2 \approx \frac{6}{\sigma}  \quad \quad , \quad \quad K=2 \ .
\end{equation}

Numerical solutions of these differential equations are greatly consistent with the general solution \eqref{fjhj} with constants \eqref{var}. Also it is a simple task to substitute the proposed solution of intermediate region into the field equations \eqref{I00-fh}-\eqref{IC} and observe that 
\begin{align}
& G_{tt}  + \alpha_q \mathcal{Q}_t = \mathcal{O}(\sigma^{-2}) \ , \nonumber\\
&  G_{rr}  + \alpha_q \mathcal{Q}_r = \mathcal{O}(\sigma^{-2}) \ , \nonumber \\
& G_{\theta\theta} + \alpha_q \mathcal{Q}_{\theta} = \mathcal{O}(\sigma^{-4}) \ , \nonumber \\
& \mathcal{Q}_c = \mathcal{O}(\sigma^{-2}) \ .
\end{align}
Noting that for an ordinary pure quantum gravastar $\sigma=r_s/\ell_p$ is very very large, and the solution is acceptable.

\begin{figure}[h]
 \centering
  \includegraphics[width=0.6\textwidth]{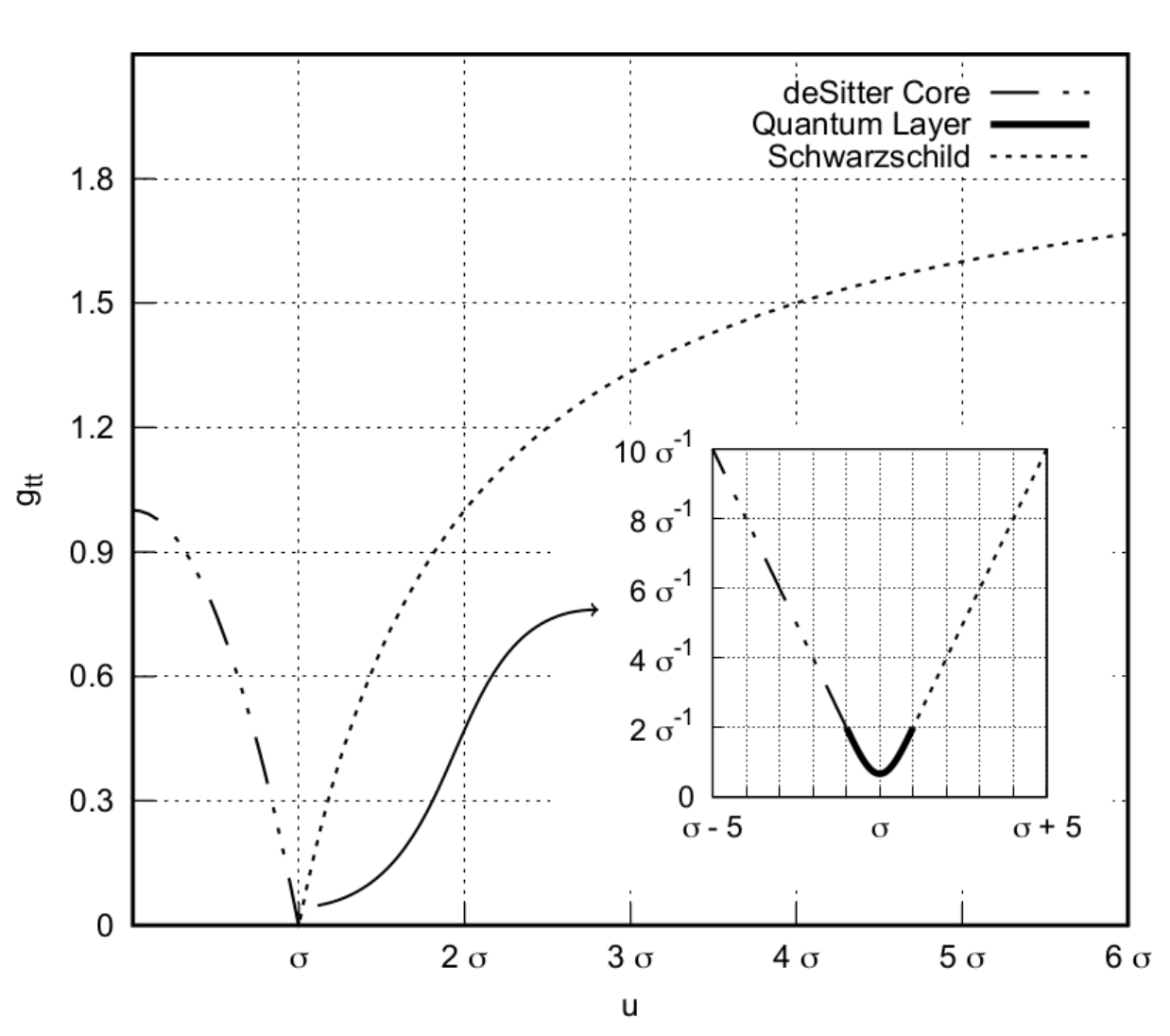}
 \caption{The temporal component of the metric as a function of radial coordinate.}
 \label{gtt}
 \end{figure}

The temporal component of the metric is plotted in figure \ref{gtt}.
Obviously, the quantum layer, extended from $-1/\sigma$ to $1/\sigma$, smoothly matches to the deSitter core at $ u =-1/\sigma$ and to the Schwarzschild region at $u=1/\sigma$. No redshift infinity surface is present and the asymptotic observer does not figure out any problem with the clock of free falling observer when crossing the quantum layer.

\begin{figure}[h]
 \centering
\includegraphics[width=0.6\textwidth, trim= 1 1 1 1,clip=true]{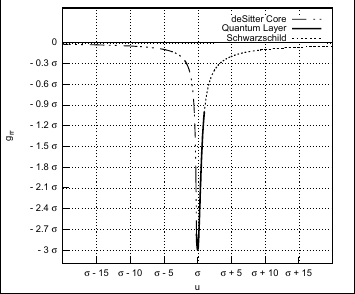}
\caption{The radial component of the metric about the quantum layer.}
 \label{grr}
\end{figure}

Also figure \ref{grr} shows the radial component of the metric. Again one has a smooth matching of the radial metric component at both deSitter and Schwarzschild boundaries and no singularity is present. Furthermore, the metric signature is saved.

As expected the divergences of the metric at the classical position of horizons are converted to a finite Gaussian curve, with height $1.5\sigma$ and its center at $\sigma$. Thus, for a big enough gravastar $(\sigma \gg 1)$, this peak behaves like a delta function. Anyway, both the horizons (of deSitter and Schwartzschild) are replaced by a quantum vacuum layer.

\section{Entropy of the quantum layer}
In order to discuss the stability of the solution, we need to assign some entropy to the quantum layer.
There is no straightforward approach to do so. This is because neither we are dealing with a matter thin layer as in the Mazur--Mottola gravastar, nor we have horizons to use surface gravity.
Let us try to define an effective energy--momentum tensor $T^{\mu (eff)}_{\nu} = (\rho_{eff},-P_{eff},-P_{eff},-P_{eff})$ for the quantum layer. Since the layer is very thin, the components of the effective energy--momentum tensor can be approximated by the average values $\overline{\alpha_q\mathcal{Q}}_t$, $\overline{\alpha_q\mathcal{Q}}_r$ and $\overline{\alpha_q\mathcal{Q}}_{\theta}$ (averaged over the volume of the layer). From equations \eqref{Iterm} to \eqref{I22-fh}, we get
\begin{align}
&\alpha_q \mathcal{Q}_t \simeq 56.9 \cdot \frac{\alpha_q}{\sigma} -3072 \ \alpha_q\sigma^2 \epsilon \ ,\\
&\alpha_q \mathcal{Q}_r \simeq -63.2 \ \alpha_q\sigma -227.6  \ \alpha_q \sigma \epsilon \ , \label{Ef-Pressure}\\
&\alpha_q \mathcal{Q}_{\theta} \simeq -6.3 \ \frac{\alpha_q}{\sigma^2} - 85.3 \ \frac{\alpha_q}{\sigma} \epsilon
\end{align}
where $\epsilon = \frac{u}{\sigma} - 1$. Considering the thin layer is defined as the region $ \sigma - 1 < u < \sigma+1 $, we have 
\begin{equation}
-\frac{1}{\sigma} < \epsilon < \frac{1}{\sigma} \ ,
\end{equation}
and the average values of $\alpha_q\mathcal{Q}_t$, $\alpha_q\mathcal{Q}_r$ and $\alpha_q\mathcal{Q}_{\theta}$ on the quantum layer are
\begin{align}
&\overline{\alpha_q \mathcal{Q}}_t = \frac{\sigma}{2} \int_{-1/\sigma}^{1/\sigma}  \alpha_q \mathcal{Q}_t  \dd{\epsilon} = 56.8 \ \frac{\alpha_q}{\sigma}  \ ,\\
&\overline{\alpha_q \mathcal{Q}}_r  = \frac{\sigma}{2} \int_{-1/\sigma}^{1/\sigma}  \alpha_q \mathcal{Q}_t  \dd{\epsilon} = -63.2 \ \alpha_q\sigma  \ ,\\
&\overline{\alpha_q \mathcal{Q}}_{\theta} = \frac{\sigma}{2} \int_{-1/\sigma}^{1/\sigma}  \alpha_q \mathcal{Q}_t  \dd{\epsilon} = -6.3 \ \frac{\alpha_q}{\sigma^2} \ \ .
\end{align}
At the limit $\sigma \gg 1$,  $\overline{\alpha_q\mathcal{Q}_t}$ and $\overline{\alpha_q\mathcal{Q}_{\theta}}$ tend to zero, and thus a well--defined equation of state parameter can not be defined for our quantum layer.

We can obtain the equivalent entropy and temperature of the quantum layer by identifying the action calculated over the layer region to the Helmholtz free energy.
Therefore, since for the quantum layer \eqref{fjhj} located at $\sigma$, the improved action is
\begin{equation} 
\mathcal{A}^{\text{I}} = \mathcal{C}\int_V \dd{V(\sigma)} \dd{(ct)} \mathcal{L}(\sigma)  
\end{equation}
where $\tilde{\alpha}_q \equiv \alpha_q / l_p$,
\begin{align}
& \mathcal{L}(\sigma) =  R(\sigma)+\epsilon_q R_{\alpha\beta}(\sigma)R^{\alpha\beta}(\sigma) = 1.6 \tilde{\alpha}_q \frac{\sigma^2}{l_p^4} + 1.8 \frac{\sigma}{l_p^4} + \mathcal{O}(\sigma^{-1}) \ , \\
&\dd{V(\sigma)} =  \sqrt{-g} \dd[3]{x} = 4\sqrt{2} \pi \sigma^3 l_p^3 \dd{\epsilon} \ ,
\end{align}
and setting $\mathcal{C}\int\dd{(ct)} = 2c/ \kappa$, we obtain the Helmholtz free energy as
\begin{equation} \label{FE}
\dd{F} \simeq -P_{eff}\dd{V}(\sigma) -  56.2 \times 4\sqrt{2}  \pi \frac{\tilde{\alpha}_q^2}{\kappa} \sigma^5 l_p^3 \dd{\epsilon}
\end{equation}
where we have used equation \eqref{Ef-Pressure} to substitute for $P_{eff}$.

On the other hand the second term in the above equation should be $SdT$. Noting the dimension of temperature, $[k_BT] \sim m_p c^2$, it can be deduced that 
\begin{equation}
\dd{k_BT} \sim m_p c^2 \dd{\epsilon} \ .
\end{equation}
Therefore, since $\kappa=8\pi l_p/ m_p c^2$ in planck units, the entropy of the quantum layer becomes
\begin{equation}
S \simeq 40 k_B \alpha_q^2 \sigma^5 \ .
\end{equation}
It is very important to note that the positivity of the entropy confirms the stability of the layer, and that larger quantum gravastars contain more entropy.

\section{Conclusion}
The gravastar was proposed by Mazur and Mottola as a stable, compact nonsingular solution of Einstein equation. In some aspects, it seems a proper alternative for the black hole as a gravitational collapse endpoint. The main features of this model are its non--sigular deSitter core and the replacment of the event horizon by a thin layer of ultra-relativistic fluid.

Here we have shown that, in the context of quantum improved Einstein equations (that is including quantum gravitational effects), it is quite possible to have a pure quantum gravastar. Such a solution consists of three regions. A deSitter core, a Schwartzschild outer region, and an intermediate region matching the other two regions for which the quantum effects are dominant.

The thin quantum layer matches smoothly with  both deSitter and Schwarzschild sides, and thus there is no need for Darmois--Israel junction layer.
This is an important result, as it would remove the need for complicated junction condition considerations for the null hypersurface. As stated originally in\cite{null-Israel}, the zero extrinsic curvature of a null hypersurface, makes difficulties in using continuity conditions of the fundamental forms. Although it can be shown that a small modification of these conditions based on the gravastar model solves the problem\cite{Mo-junction, rotating-gra}, but for this quantized one there is no need for special considerations. Unlike a classical gravastar, the quantum corrections of the geometrical vacuum at the junction location, does not lead to some surface stress tensor. Therefore, we can define a pure quantum gravastar as a region of cosmological constant (maybe dark energy) confined by quantum gravitational effects, and thus is a vacuum solution of quantum improved Einstein equations.

Since the question of the stability of such a solution is an important problem, we have studied the entropy of the obtained solution and shown that it is stable. This is done using a Helmholtz free energy, as there is no well--defined equation of state for the quantized vacuum layer.

\appendix
\renewcommand{\thesection}{Appendix:}
\section{Dominated quantum fluctuation}
To determine the most impressive term in $f(\chi)$, each term of $\chi^2 =\varphi_1 / R + \varphi_2 \times R/R_{\alpha\beta}R^{\alpha\beta} + \varphi_3 \times R/ R_{\alpha\beta\gamma\delta}R^{\alpha\beta\gamma\delta}$ should be evaluated about the horizon separately. Since the cutoff momentum $k^2 = \xi_0 \chi^{-2}$ is more intuitive in the context of the fluctuation, we proceed with it instead of $\chi^2$.

The vacuum solution \eqref{fjhj} leads to
\begin{align}
& \abs{R} = \frac{1}{r_s^2}\frac{2(16\sigma^2 - 9 \sigma +6)}{ 9 \sigma} \ ,\\
& \abs{\frac{R_{\alpha\beta}R^{\alpha\beta}}{R}} = \frac{1}{2} \abs{\frac{R_{\alpha\beta\gamma\delta}R^{\alpha\beta\gamma\delta}}{R}}= \frac{1}{r_s^2} \frac{256\sigma^4+81\sigma^2-108\sigma+36}{9\sigma (16\sigma^2 - 9 \sigma +6)} \ .
\end{align}
Clearly, for our improved spherical static vacuum solution all these three terms are of the same order, and none of them can dominate the other two at least up to the first order of perturbation. Since we are looking for the possibility of such a model, restrict ourselves to $\chi^2 = \varphi_2 \times R/R_{\alpha\beta}R^{\alpha\beta}$, which reduces the complexity of the improved field equations and makes the comparison of the results with other improved black hole models\cite{our2} possible.

At the end, it should be noted that a straightforward calculation shows that the same layer \eqref{fjhj} is derived if we choose other suggested curvature invariants, i.e.  $1 / R$ and $ R/ R_{\alpha\beta\gamma\delta}R^{\alpha\beta\gamma\delta}$ for cutoff identification up to the first order of perturbation. This is an invaluable point, since, as it is stated before, the process of choosing between these three is a debatable issue in the improvement method\cite{our3,Reuter-2021}.

\end{document}